\documentclass[12pt]{article}
\usepackage{comment}
\usepackage{amsmath}
\usepackage{amssymb}
\usepackage{graphicx}
\usepackage{here}
\usepackage{subcaption}
\usepackage{url}
\usepackage[sort&compress, numbers, merge]{natbib}
\usepackage{braket}
\usepackage{physics}
\usepackage[compat=1.1.0]{tikz-feynhand}
\usepackage{slashed}
\usepackage{mathtools}
\usepackage{booktabs}
\usepackage{multirow}

\usepackage{todonotes}

\numberwithin{equation}{section}

\usepackage[colorlinks=true, linkcolor=blue, citecolor=blue,
urlcolor=black]{hyperref}

\begin{document}
\def\ps{\mathbf{p}}
\def\PS{\mathbf{P}}
\baselineskip 0.6cm
\def\simgt{\mathrel{\lower2.5pt\vbox{\lineskip=0pt\baselineskip=0pt
           \hbox{$>$}\hbox{$\sim$}}}}
\def\simlt{\mathrel{\lower2.5pt\vbox{\lineskip=0pt\baselineskip=0pt
           \hbox{$<$}\hbox{$\sim$}}}}
\def\simprop{\mathrel{\lower3.0pt\vbox{\lineskip=1.0pt\baselineskip=0pt
             \hbox{$\propto$}\hbox{$\sim$}}}}
\def\tr{\mathop{\rm tr}}
\def\SU{\mathop{\rm SU}}
\def\Umt{\mathrm{U}(1)_{L_{\mu}-L_{\tau}}}
\def\calL{\mathcal{L}}
\def\calM{\mathcal{M}}
\def\Nbar{\overline{N}}
\def\suthree{\mathrm{SU}(3)}
\def\sutwo{\mathrm{SU}(2)}
\def\uone{\mathrm{U}(1)}
\def\ubar{\overline{u}}
\def\dbar{\overline{d}}
\def\ebar{\overline{e}}
\def\s12{s_{12}}

\begin{titlepage}

\begin{flushright}
IPMU25-0011
\end{flushright}

\vskip 1.1cm

\begin{center}

{\Large \bf 
Global Neutrino Constraints on the Minimal $\boldsymbol{\Umt}$ Model 
}

\vskip 1.2cm
Masahiro Ibe$^{a,b}$, 
Satoshi Shirai$^{b}$ and
Keiichi Watanabe$^{a}$
\vskip 0.5cm

{\it

$^a$ {ICRR, The University of Tokyo, Kashiwa, Chiba 277-8582, Japan}

$^b$ {Kavli Institute for the Physics and Mathematics of the Universe
(WPI), \\The University of Tokyo Institutes for Advanced Study, \\ The
University of Tokyo, Kashiwa 277-8583, Japan}

}

\vskip 1.0cm

\abstract{
We examine the minimal $\Umt$ gauge model in light of the latest neutrino data, 
including neutrino oscillations, 
cosmological observations, 
direct mass measurements, 
and neutrinoless double-beta decay.  
Using the most conservative oscillation data,  
we find that normal ordering is excluded at approximately the 90\% confidence level (CL).  
Incorporating cosmological constraints from Cosmic Microwave Background (CMB) observations strengthens this exclusion to about 95\%\,CL,  
while further including Baryon Acoustic Oscillation (BAO) data increases it to nearly 99\%\,CL.  
The inverted ordering is even more strongly disfavored.  
Our analysis is performed within a frequentist framework,  
minimizing sensitivity to prior assumptions inherent in Bayesian approaches.  
These results impose strong constraints on the viability of the minimal $\Umt$ gauge model.
}

\end{center}
\end{titlepage}

\section{Introduction} 
The $\uone_{L_\mu - L_\tau}$ gauge symmetry\,\cite{Foot:1990mn, He:1991qd, Foot:1994vd, Gninenko:2001hx, Baek:2001kca, Murakami:2001cs, Ma:2001md} has long been an intriguing subject of study within the particle phenomenology community as one of the simplest extensions of the Standard Model (SM) that remains consistent with experimental constraints. 
In particular, 
the minimal $\uone_{L_\mu - L_\tau}$ model, 
where the symmetry is spontaneously broken by the vacuum expectation value of a single $\uone_{L_\mu - L_\tau}$ charged scalar field, 
realizes the two-zero minor neutrino 
mass structure\,\cite{Lavoura:2004tu,Lashin:2007dm}. 
This leads to high predictive power for neutrino oscillation phenomena, 
which has therefore gathered significant attention in the context of neutrino physics.

Furthermore, the model has potential implications for the anomalous magnetic moment of the muon ($g-2$) 
when the gauge boson mass of $\Umt$ lies in the sub-GeV region. 
At present, it remains uncertain whether a discrepancy exists between the experimental measurements\,\cite{Muong-2:2006rrc,Muong-2:2021ojo,Muong-2:2021vma} of the muon’s anomalous magnetic moment and the SM predictions\,\cite{Aoyama:2020ynm}.%
\footnote{%
The theoretical prediction of the muon anomalous magnetic moment ($g-2$) remains somewhat unclear at present. 
Currently, discrepancies between data-driven approaches and lattice simulations have been reported 
(see Ref.\,\cite{ParticleDataGroup:2024cfk} and references therein). 
} 
However, if such a discrepancy is confirmed, 
the $\Umt$ symmetry is widely regarded as a well-motivated extension that could account for this anomaly. 
Various experimental searches for this model are actively underway (see e.g., Refs.\,\cite{Harnik:2012ni,Bilmis:2015lja,BaBar:2016sci,Ibe:2016dir,Capdevilla:2021kcf,Belle-II:2023ydz,Belle-II:2024wtd,NA64:2024nwj}).

On the other hand, 
as pointed out in Refs.\,\cite{Asai:2017ryy,Asai:2018ocx,Asai:2020qax}, 
the high predictive power of the minimal $\Umt$ model strongly constrains it through the results of neutrino oscillation experiments and cosmological observations, 
particularly constraints on the sum of neutrino masses. 
Consequently, 
analyses based on the latest data suggest that this model is increasingly entering a region in conflict with observational results.

In this paper,  
we investigate the extent to which the minimal $\Umt$ model is statistically disfavored based on the latest data from neutrino oscillation experiments,  
direct mass measurements, 
neutrinoless double-beta decay,  
Cosmic Microwave Background (CMB) observations,  
and Baryon Acoustic Oscillation (BAO) observations.  
Our analysis shows that,  
under the assumption of the $\Lambda$CDM model,  
the minimal $\Umt$ model is excluded at the 99\% confidence level (CL).  
This result suggests that reviving the minimal $\Umt$ model would require significant modifications beyond the $\Lambda$CDM framework. 
It is important to note that our results apply specifically to the minimal $\Umt$ model and do not exclude the possibility of more complex extensions.

The organization of this paper is as follows.
In Sec.\,\ref{sec:model}, 
we summarize a minimal gauged $\Umt$ model. 
In Sec.\,\ref{sec:analysis},
we analyze this model from the perspective of frequentist statistics.
The final section is devoted to our conclusions.

\section{Minimal \texorpdfstring{$\boldsymbol{\Umt}$}{} Model}
\label{sec:model}

\begin{table}[t!]
\begin{center}
\renewcommand{\arraystretch}{1.5}
\caption{Charge assignments of leptons and $\phi$ under $\Umt$ gauge symmetry. All other SM fields not listed in this table are neutral under this symmetry.
}
\begin{tabular}{ccc} \hline
   \multicolumn{2}{c}{Field} & $\Umt$ \\ \hline
   \multirow{3}{*}{Leptons} 
    & $L_{e}$ \,\,\, $\ebar_{e}$ \,\,\, $\Nbar_{e}$ & $0$ \\
   & $L_{\mu}$ \,\,\, $\ebar_{\tau}$ \,\,\, $\Nbar_{\tau}$ & $+1$ \\
    & $L_{\tau}$ \,\,\, $\ebar_{\mu}$ \,\,\, $\Nbar_{\mu}$ & $-1$ \\ \hline
   \multirow{1}{*}{Scalar} & $\phi$ & $+1$ \\ \hline
\end{tabular}
\label{tab:mutau_charge}
\end{center}
\end{table}

We summarize the minimal gauged $\Umt$ model in Refs.\,\cite{Asai:2017ryy,Asai:2018ocx,Asai:2020qax}.
The model extends the SM by introducing three right-handed neutrinos
$\Nbar_{e,\mu,\tau}$ and a complex scalar $\phi$.%
\footnote{In this paper, 
all the fermions are denoted by the Weyl fermion with the notation used in Ref.\,\cite{Dreiner:2008tw}.
}
This model has a new $\uone$ symmetry under which leptons and $\phi$ have $\uone$ charges shown in Tab.\,\ref{tab:mutau_charge}.
All other particles not listed in the table do not have this $\Umt$ charge.

Note that in a renormalizable model where there is only one complex scalar field that spontaneously breaks $\Umt$ symmetry, 
the (type-I) seesaw mechanism\,\cite{Minkowski:1977sc,Yanagida:1979as,*Yanagida:1979gs,Gell-Mann:1979vob,Glashow:1979nm,Mohapatra:1979ia} is the minimal setup 
to reproduce the neutrino oscillation experiments.%
\footnote{
For other $\Umt$ models based on alternative seesaw mechanisms, see the Appendix.}
With the gauge charges given in Tab.\,\ref{tab:mutau_charge}, the interaction terms responsible for neutrino mass and mixing are given by, 
\begin{align}
\calL
= - \sum_{i=e,\mu,\tau} \kappa_i\, L_i\, \Nbar_i\, H 
- \frac{1}{2} M_{ee} \Nbar_e\, \Nbar_e - M_{\mu \tau}  \Nbar_\mu\, \Nbar_\tau 
- \kappa_{e \mu} \Nbar_e\, \Nbar_\mu\, \phi
- \kappa_{e \tau} \Nbar_e\, \Nbar_\tau\, \phi^{\dagger} + \mathrm{h.c.}
\end{align}
Here, 
$H$ denotes the SM Higgs doublet. 
Due to the $\Umt$ gauge symmetry, 
the Yukawa interactions of the charged leptons are diagonal, $\kappa_{e,\mu,\tau}$. 
Moreover, 
by exploiting the phase freedom of the singlet charged leptons, these Yukawa couplings can be made real and positive without loss of generality.
Similarly, 
the Yukawa interactions $\kappa_i$ between the right-handed neutrinos and the lepton doublets are also diagonal, 
and can be rendered real and positive by utilizing the phase freedom of the lepton doublets.
Additionally, 
the mass terms $M_{ee}$ and $M_{\mu\tau}$ can be made real and positive through the phase rotations of the right-handed neutrinos. However, 
it is not possible to simultaneously render both $\kappa_{e\mu}$ and $\kappa_{e\tau}$ real-valued using phase transformations.
Consequently, 
the minimal $\Umt$ model inherently contains a single physical parameter responsible for CP violation.
We will make this point more explicit later by employing the constraint equations associated with the two-zero minor structures.

After the $H$ and $\phi$ acquire their vacuum expectation values (VEVs), 
$\langle H \rangle = (0,v_\mathrm{EW})^T$ and $\langle \phi \rangle = v_\phi$, respectively,
the following mass terms arise:
\begin{align}
\calL
= - m_{D,ij}\, L_i\, \Nbar_i
- \frac{1}{2}\, M_{R,ij}\, \Nbar_i\, \Nbar_j 
+ \mathrm{h.c.}\ ,
\end{align}
where $m_D$ and $M_{R}$ are expressed as
\begin{align}
m_D
= 
v_{\mathrm{EW}}
\begin{pmatrix}
\kappa_e & 0 & 0 \\
0 & \kappa_\mu & 0 \\
0 & 0 & \kappa_\tau
\end{pmatrix}\ , \quad 
M_R 
= 
\begin{pmatrix}
M_{ee} & \kappa_{e \mu} v_\phi & \kappa_{e \tau} v_\phi \\
\kappa_{e \mu} v_\phi & 0 & M_{\mu \tau} \\
\kappa_{e \tau} v_\phi & M_{\mu \tau} & 0
\end{pmatrix}\ .
\end{align}
By integrating out the right-handed neutrinos,
the mass matrix of light neutrino are given by the seesaw mechanism\,\cite{Minkowski:1977sc,Yanagida:1979as,*Yanagida:1979gs,Gell-Mann:1979vob,Glashow:1979nm,Mohapatra:1979ia},
\begin{align}
\label{eq:seesaw_mass}
m_{\nu} = - m_D\, M_R^{-1}\, m_D^T\ .
\end{align}
This matrix is a complex symmetric matrix and 
can be diagonalized using the Takagi decomposition with a single unitary matrix,
\begin{align}
m_{\nu} 
= U^{*} \, \hat{m}_{\nu} \, 
U^{\dagger}\ , 
\quad \hat{m}_{\nu} = \mathrm{diag}(m_1, m_2, m_3)\ ,
\end{align}
where $m_{1,2,3}$ are taken real positive.
The neutrino mass parameters are 
ordered
\begin{align}
    m_{1}<m_{2}<m_{3}\ ,
\end{align}
for the normal ordering (NO) case,
and 
\begin{align}
    m_{3}<m_{1}<m_{2}\ ,
\end{align}
for the inverted ordering (IO) case.
As the charged lepton sector has been diagonalized, the unitary matrix $U$ above is identified as the PMNS matrix with Majorana phases, 
and its parametrization is given by the following expression,
\begin{align}
U
&= 
\begin{pmatrix}
U_{e1}&U_{e2}&U_{e3}\\
U_{\mu 1}&U_{\mu 2}&U_{\mu 3}\\
U_{\tau 1}&U_{\tau 2}&U_{\tau 3}
\end{pmatrix} \nonumber \\
&= 
\begin{pmatrix}
c_{12}c_{13} & s_{12} c_{13} & s_{13} e^{-i \delta_{\mathrm{CP}}} \\
- s_{12} c_{23} - c_{12} s_{13} s_{23} e^{i \delta_{\mathrm{CP}}} 
& c_{12} c_{23} - s_{12} s_{13} s_{23} e^{i \delta_{\mathrm{CP}}} 
& c_{13} s_{23} \\
s_{12} s_{23} - c_{12} s_{13} c_{23} e^{i \delta_{\mathrm{CP}}} 
& - c_{12} s_{23} - s_{12} s_{13} c_{23} e^{i \delta_{\mathrm{CP}}} 
& c_{13} c_{23}
\end{pmatrix}
\begin{pmatrix}
e^{i \eta_1} & 0 & 0 \\
0 & e^{i \eta_2} & 0 \\
0 & 0 & 1
\end{pmatrix}\ ,
\end{align}
where $c_{ij} := \mathrm{cos}\, \theta_{ij}$ and $s_{ij} := \mathrm{sin}\, \theta_{ij}$.
Phases $\delta_{\mathrm{CP}}$ and  $\eta_{1,2}$ are the Dirac phase and the Majorana phases.

Due to the two-zero texture of $M_R$ and Eq.\,\eqref{eq:seesaw_mass},
$m_\nu^{-1}$ has two zeros 
in the $(\mu,\mu)$ and $(\tau,\tau)$ entries.
Those two-zero minor structures lead to the following constraints\,\cite{Verma:2011kz,Liao:2013saa,Asai:2017ryy},
\begin{align}
&r_{21}:=\frac{m_2}{m_1}=
    \qty(1 - \frac{\csc\theta_{12}}{\sin\theta_{12} + e^{i \delta_\mathrm{CP}} \cos\theta_{12} \sin\theta_{13} \tan 2 \theta_{23}})
    \times e^{-2i(\eta_1-\eta_2)}
    \ ,
    \label{eq:r21}
    \\
&r_{31}:=\frac{m_3}{m_1}
= 
\frac
{-\cos^2 \theta_{13}
\left[
2\cos^2\theta_{12}\cos 2\theta_{23} -\sin 2\theta_{12} \sin
 2\theta_{23} \sin \theta_{13} e^{i\delta_\mathrm{CP}}\right]}
{\sin\theta_{13} \left[
2\cos 2\theta_{12} \cos 2\theta_{23} \sin \theta_{13} 
-\sin 2\theta_{12} \sin 2\theta_{23} (e^{-i\delta_\mathrm{CP}}+\sin^2 \theta_{13} e^{i\delta_\mathrm{CP}})
\right]
}
\times e^{-i(2\delta_\mathrm{CP}+2\eta_1)}
\ .
\label{eq:r31}
\end{align}
The quantities $r_{21}$ and $r_{31}$ on the left-hand side of Eqs.,\eqref{eq:r21} and \eqref{eq:r31} are positive real numbers, while the right-hand side is generally complex.
To ensure that the right-hand side also becomes a positive real number, two of the three CP phases $\eta_1$, $\eta_2$, and $\delta_\mathrm{CP}$ must be fixed in terms of the remaining one.
In our analysis, we take $\delta_\mathrm{CP}$ as the free parameter and determine $\eta_1$ and $\eta_2$ accordingly.

\begin{figure}[t!]
	\centering
{\includegraphics[width=0.4\textwidth]{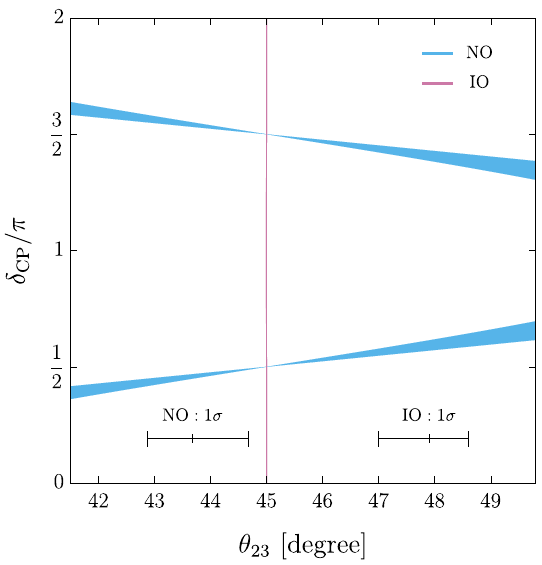}}
\hspace{1cm}
{\includegraphics[width=0.426\textwidth]{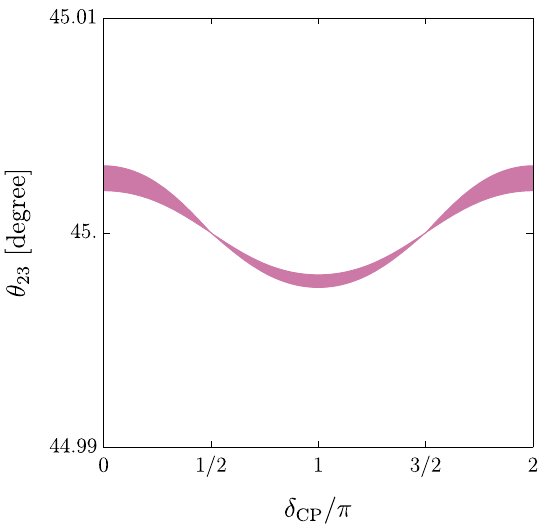}}

 \caption{The correlation between $\delta_\mathrm{CP}$ 
 and the mixing parameters
 in the minimal $\Umt$ model.
The blue lines show $\delta_\mathrm{CP}$ as a function of $\theta_{23}$ for the NO case.
In the left figure, the purple vertical line indicates the predicted value $\theta_{23} \simeq \pi/4$ for the IO case, which is closed up in the right panel.
Mixing parameters and  $R_{\mathrm{NO,IO}}$ are varied within their 
$3\sigma$ ranges according to NuFIT~$6.0$ (\texttt{IC24 with SK-atm}),
where
the $1\sigma$ ranges of $\theta_{23}$ are indicated by bars.
}
 \label{fig:params}
\end{figure}

To see the correlation between parameters, 
let us consider the ratio between the solar and atmospheric mass-squared differences,
\begin{align}
    R_{\mathrm{NO}} := \frac{\Delta m_{21}^2}{\Delta m_{31}^2} 
     = \frac{r_{21}^2-1}{r_{31}^2 - 1} \ , 
\end{align}
for the NO case and 
 \begin{align}
    R_{\mathrm{IO}} := \frac{\Delta m_{12}^2}{\Delta m_{32}^2} 
     = \frac{1-r_{21}^2}{r_{31}^2 - r_{21}^2} \ , 
\end{align}
for the IO case.
Then, 
by taking the observed values of $R_{\mathrm{NO,IO}}$ extracted from NuFIT~$6.0$\,\cite{Esteban:2024eli}, 
we show the correlation 
between $\theta_{23}$
and $\delta_\mathrm{CP}$ in Fig.\,\ref{fig:params}.
Other parameters and 
the value of 
$R $ are varied within their $3\sigma$ ranges according to NuFIT~$6.0$ (\texttt{IC24 with SK-atm}),
where
the $1\sigma$ ranges
of $\theta_{23}$ are indicated by bars.

For the NO case,  
the correlation between the mixing angles and $\delta_\mathrm{CP}$ can be understood as follows. 
From Eq.\,\eqref{eq:r31}, we find that $r_{31}$ is at most $\order{1}$.  
Thus, the observed value $R_{\mathrm{NO}} = \order{10^{-2}}$ is realized only when $r_{21} \simeq 1$,  
which leads to  
\begin{align}
\cos \delta_\mathrm{CP} \simeq \cot 2\theta_{12} \cot 2\theta_{23} \csc \theta_{13} \ .
\end{align}
Using the best-fit values of the mixing angles from NuFIT 6.0,  
we obtain $\cos\delta_\mathrm{CP} \simeq 0.17$,  
corresponding to $\delta_\mathrm{CP}$ values around $\pi/2$ and $3\pi/2$ for the NO case.  

For the IO case,  
on the other hand,  
the correlation emerges along $\theta_{23} \simeq \pi/4$,  
where $r_{21} \simeq 1$.  
Expanding $s_{23}$ around $\theta_{23} \simeq \pi/4$ as  
$s_{23} \simeq 1/\sqrt{2} + \epsilon$ ($|\epsilon| \ll 1$),  
we find  
\begin{align}
r_{21} \simeq 1 + \frac{4\sqrt{2}\cos\delta_\mathrm{CP}}{\sin\theta_{13} \sin 2\theta_{12}}\, \epsilon + \order{\epsilon^2} \ .
\end{align}
Thus, the observed value $R_{\mathrm{IO}} = \order{10^{-2}}$ can be achieved with a small $\epsilon$.

\begin{figure}[t!]
	\centering
{\includegraphics[width=0.4\textwidth]{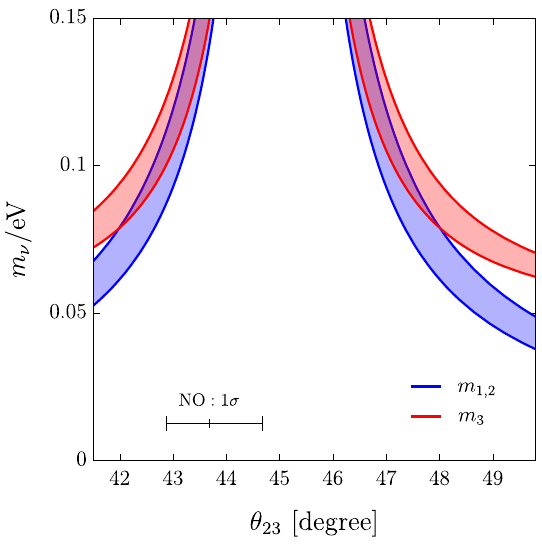}}
\hspace{1cm}
{\includegraphics[width=0.4\textwidth]{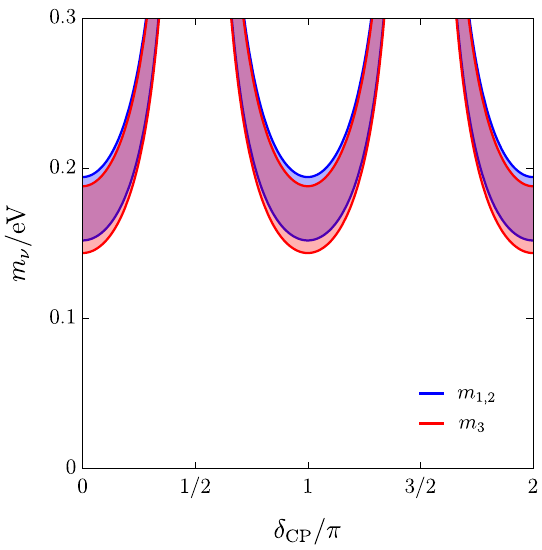}}
 \caption{
The neutrino mass spectrum as a function of $\theta_{23}$ for the NO case (left) and $\delta_\mathrm{CP}$ for the IO case (right), varying the mixing parameters within their $3\sigma$ ranges based on NuFIT~6.0 (\texttt{IC24 with SK-atm}).
For the IO case, $\theta_{23}$ is in the vicinity of $\pi/4$.
}
 \label{fig:total_mass}
\end{figure}

For both the NO and IO cases, 
we find that $r_{21} \simeq 1$, 
implying  
\begin{align}
m_{1}^2\ , \, m_{2}^2  \gg \Delta m_{21}^2 \ ,
\end{align}
which indicates that $m_1$ and $m_2$ are nearly degenerate.  
Figure\,\,\ref{fig:total_mass} presents the neutrino mass spectrum as a function of $\theta_{23}$ for the NO case and $\delta_\mathrm{CP}$ for the IO case,  
with the oscillation parameters varied within their $3\sigma$ ranges based on NuFIT~6.0 (\texttt{IC24 with SK-atm}).
For the NO case, 
the spectrum exhibits mild degeneracy between $m_3$ and $m_{1,2}$,  
whereas in the IO case, the masses are fully degenerate, particularly near $\theta_{23} \simeq \pi/4$.  
This strong degeneracy results in a relatively large total neutrino mass in the minimal $\Umt$ model.  
As discussed in the next section, cosmological observations impose stringent constraints on the total neutrino mass,  
placing significant restrictions on the model's viability.

Before closing this section,  
we briefly comment on the radiative corrections to the two-zero minor structure of the neutrino mass matrix.  
At the leading order, this structure remains invariant under renormalization group running below the seesaw scale.  
However, 
at the one-loop level, 
finite threshold corrections from the $\Umt$ gauge symmetry breaking sector induce right-handed neutrino mass terms,  
$M_{\mu\mu}$ and $M_{\tau\tau}$, proportional to $\kappa_{e\mu}^2 \langle \phi \rangle^2$ and  
$\kappa_{e\tau}^2 \langle \phi^\dagger \rangle^2$, respectively.  
As a consequence, the two-zero minor structure is explicitly broken at one loop.  
In this study, we assume that the couplings between the $\Umt$-breaking field and the right-handed neutrinos are sufficiently small,  
allowing us to neglect these corrections.

\section{Analysis and Result}
\label{sec:analysis}

Neutrino physics is characterized by nine parameters: the three neutrino masses, three mixing angles, and three CP phases.  
In the $\Umt$ model, imposing the two-zero minor condition leads to the constraints in Eqs.\,\eqref{eq:r21} and \eqref{eq:r31},
which reduces the number of free parameters by four, leaving a total of five.  
To constrain the $\Umt$ model, one can apply a likelihood ratio test based on Wilks' theorem,  
comparing the full neutrino model with its $\Umt$-restricted counterpart to derive parameter constraints.  

Tables\,\,\ref{tab:NO_summary} and \ref{tab:IO_summary} summarize the results.  
The first row indicates the type of observational data used in the analysis, along with $\Delta \chi^2$ and the exclusion confidence level (CL).  
Here, $\Delta \chi^2$ is defined as the difference between the $\chi^2$ minima of the restricted model (minimal $\Umt$ model) and the full model.  
From the second row onward, the table specifies the data included in each analysis,  
where ``-'' denotes that the corresponding dataset was not used.  

Following the convention of NuFIT, we present two sets of results for the IceCube experiment and Super-Kamiokande atmospheric neutrino data:  
``\texttt{IC19 w/o SK-atm}'' and ``\texttt{IC24 with SK-atm}.''  
A dagger ($\dagger$) next to the exclusion confidence level indicates that the value is determined solely by neutrino oscillation data.  
The details of the analysis will be discussed later.

The tables show that, using the most conservative oscillation data,  
NO is excluded at approximately the 90\%\,CL.  
Incorporating cosmological constraints from the Planck~2020~PR4 analysis strengthens this exclusion to about 95\%\,CL,  
while further including BAO data increases it to nearly 99\%\,CL.  
For the IO case, 
even with the most conservative neutrino oscillation data,  
the model is excluded at the 92\%\,CL.  
With the addition of cosmological constraints, even the most conservative limits from Planck~2020~PR4  
exclude the model at a significance of $4.4\sigma$.

\begin{table}[t]
    \centering
\caption{Constraints on the NO case using NuFIT 6.0 oscillation data. 
The results are presented separately for analyses \texttt{IC19 w/o SK-atm} and 
\texttt{IC24 with SK-atm}. 
Each row includes additional constraints from direct neutrino mass measurements ($m_{\beta}$), neutrinoless double beta decay ($0\nu\beta\beta$) with different nuclear matrix elements ($M^{0\nu}$), and cosmological observations (Planck, Planck + DESI). The confidence levels (CL) or statistical significance ($\sigma$) are reported for each case.}
\label{tab:NO_summary}
    \begin{tabular}{c|c|c|c|c|c} \hline
        Oscillation & $m_{\beta}$ & $0\nu\beta\beta$  & Cosmology & $\Delta \chi^2$(NO) & Confidence Level \\ \hline
        \multicolumn{6}{c}{\textbf{IC19 w/o SK-atm (NuFIT 6.0)}} \\ \hline
        NuFIT 6.0 & - & - & - & $4.3$ & $96\%$\,CL$^{\phantom{\dagger}}$ \\
        NuFIT 6.0 & KATRIN & - & - & $4.4$ & $96\%$\,CL$^\dagger$ \\ \hline
        NuFIT 6.0 & - & $M^{0\nu} = 1.11$ & - & $5.1$ & $96\%$\,CL$^\dagger$ \\
        NuFIT 6.0 & - & $M^{0\nu} = 2.39$ & - & $7.6$ & $96\%$\,CL$^\dagger$ \\
        NuFIT 6.0 & - & $M^{0\nu} = 4.77$ & - & $10$ & $97\%$\,CL$^{\phantom{\dagger}}$ \\ \hline
        NuFIT 6.0 & - & - & Planck & $6.6$ & $96\%$\,CL$^{\phantom{\dagger}}$ \\
        NuFIT 6.0 & - & - & Planck + DESI & $13$ & $3.2\sigma$  \\ \hline
        \multicolumn{6}{c}{\textbf{IC24 with SK-atm (NuFIT 6.0)}} \\ \hline
        NuFIT 6.0 & - & - & - & $2.7$ & $90\%$\,CL$^{\phantom{\dagger}}$ \\
        NuFIT 6.0 & KATRIN & - & - & $2.9$ & $90\%$\,CL$^\dagger$ \\ \hline
        NuFIT 6.0 & - & $M^{0\nu} = 1.11$ & - & $3.8$ &  $90\%$\,CL$^\dagger$  \\
        NuFIT 6.0 & - & $M^{0\nu} = 2.39$ & - & $7.4$ &  $90\%$\,CL$^\dagger$ \\
        NuFIT 6.0 & - & $M^{0\nu} = 4.77$ & - & $15$ & $2.8\sigma$ \\ \hline
        NuFIT 6.0 & - & - & Planck & $5.8$ & $94\%$\,CL$^{\phantom{\dagger}}$ \\
        NuFIT 6.0 & - & - & Planck + DESI & $16$ & $3.6\sigma$  \\ \hline
    \end{tabular}
\end{table}

\begin{table}[t]
    \centering
    \caption{Same as Tab.\,\ref{tab:NO_summary}, but for the IO case.}
    \label{tab:IO_summary}
    \begin{tabular}{c|c|c|c|c|c} \hline
        Oscillation & $m_{\beta}$ & $0\nu\beta\beta$ & Cosmology & $\Delta \chi^2$(IO) & Confidence Level \\ \hline
        \multicolumn{6}{c}{\textbf{IC19 w/o SK-atm (NuFIT 6.0)}} \\ \hline
        NuFIT 6.0 & - & - & - & $5.8$ & $98\%$\,CL \\
        NuFIT 6.0 & KATRIN & - & - & $9.1$ & $2.6\sigma$ \\ \hline
        NuFIT 6.0 & - & $M^{0\nu} = 1.11$ & - & $20$ & $3.5\sigma$ \\
        NuFIT 6.0 & - & $M^{0\nu} = 2.39$ & - & $65$ & $7.3\sigma$ \\ \hline
        NuFIT 6.0 & - & - & Planck & $24$ & $4.6\sigma$ \\
        NuFIT 6.0 & - & - & Planck + DESI & $68$ & $8.1\sigma$ \\ \hline
        \multicolumn{6}{c}{\textbf{IC24 with SK-atm (NuFIT 6.0)}} \\ \hline
        NuFIT 6.0 & - & - & - & $3.1$ & $92\%$\,CL \\
        NuFIT 6.0 & KATRIN & - & - & $7.1$ & $97\%$\,CL \\ \hline
        NuFIT 6.0 & - & $M^{0\nu} = 1.11$ & - & $18$ & $3.2\sigma$  \\
        NuFIT 6.0 & - & $M^{0\nu} = 2.39$ & - & $63$ & $7.2\sigma$  \\ \hline
        NuFIT 6.0 & - & - & Planck & $23$ & $4.4\sigma$  \\
        NuFIT 6.0 & - & - & Planck + DESI & $68$ & $8.0\sigma$  \\ \hline
    \end{tabular}
\end{table}

\subsection{Neutrino oscillation}

First, we conduct a likelihood analysis of the minimal $\Umt$ model based on neutrino oscillation experiments using data from NuFIT~6.0\,\cite{Esteban:2024eli}.
NuFIT provides two analyses: 
``\texttt{IC19 w/o SK-atm},'' 
which utilizes 
IceCube/DeepCore (IC) 
3--year data (2012--2015)\,\cite{IceCube:2019dqi,IceCube:2019dyb}, 
and 
``\texttt{IC24 with SK-atm},'' 
which incorporates 
the $\chi^2$ map of 
IC 9.3--year data (2012--2021)\,\cite{IceCube:2024xjj} 
along with 
the $\chi^2$ map of 
484.2 kiloton-year Super-Kamiokande atmospheric data\,\cite{Super-Kamiokande:2023ahc}.

Neutrino oscillations are parameterized by six key variables:  
\begin{gather}
\Delta m_{21}^2\ , \quad 
\Delta m_{3\ell}^2 \ ,  \\
\sin^2\theta_{12} \ , \quad 
\sin^2\theta_{13}\ , \quad \sin^2\theta_{23}\ , \quad \delta_{\mathrm{CP}}\ .
\end{gather}
Although NuFIT~6.0 provides comprehensive $\chi^2$ data for these parameters,  
it does not include the full six-dimensional $\chi^2$ function.  
Instead, it offers $\chi^2$ tables marginalized over subsets of three, two, or one variable(s),  
denoted as $\chi_{\mathrm{3D}}$, $\chi_{\mathrm{2D}}$, and $\chi_{\mathrm{1D}}$, respectively,  
where all other parameters are optimized to minimize $\chi^2$.  
These marginalized $\chi^2$ values satisfy the relation  
\begin{align} 
\chi^2_{\mathrm{full}}(\Delta m_{21}^2, \Delta m_{3\ell}^2, \sin^2\theta_{12}, \sin^2\theta_{13}, \sin^2\theta_{23}, \delta_{\mathrm{CP}}) 
&\geq \chi^2_{\mathrm{3D}}(x, y, z)  
\geq \chi^2_{\mathrm{2D}}(x, y) 
\geq \chi^2_{\mathrm{1D}}(x)\ ,  
\end{align} 
where $x$, $y$, and $z$ represent the oscillation parameters.  
It is important to note that $\chi^2_\mathrm{full}$ is 
not sensitive to the absolute neutrino mass nor two Majorana phases.

To ensure a conservative treatment of model constraints, 
we define the effective $\chi^2$, 
$\chi^2_\mathrm{eff}$, used in our analysis as: 
\begin{align} 
\chi_\mathrm{eff}^2(\Delta m_{21}^2, \Delta m_{3\ell}^2, \sin^2\theta_{12}, \sin^2\theta_{13}, \sin^2\theta_{23}, \delta_{\mathrm{CP}}) 
= &\max\left[{\chi^2_{\mathrm{3D}}( \Delta m_{3\ell}^2, \sin^2\theta_{23},\delta_{\mathrm{CP}}) , 
\chi^2_{\mathrm{2D}}(x,y) ,\chi^2_{\mathrm{1D}}(x)}\right] \ .
\end{align}
This definition ensures that $\chi_\mathrm{eff}^2$ is always equal to or smaller than the full $\chi^2$, $\chi^2_{\mathrm{full}}$,
providing a conservative basis for evaluating model constraints.
In the subsequent analysis of neutrino oscillations, 
we perform a maximum likelihood estimation (i.e., the least $\chi^2$) using $\chi_\mathrm{eff}^2$, 
which is projected onto the total neutrino mass:
\begin{align}
    m_{\mathrm{tot}} = m_1 + m_2 + m_3 \ .
\end{align}

\begin{figure}[t!]
	\centering
 	\subcaptionbox{NO}
{\includegraphics[width=0.49\textwidth]{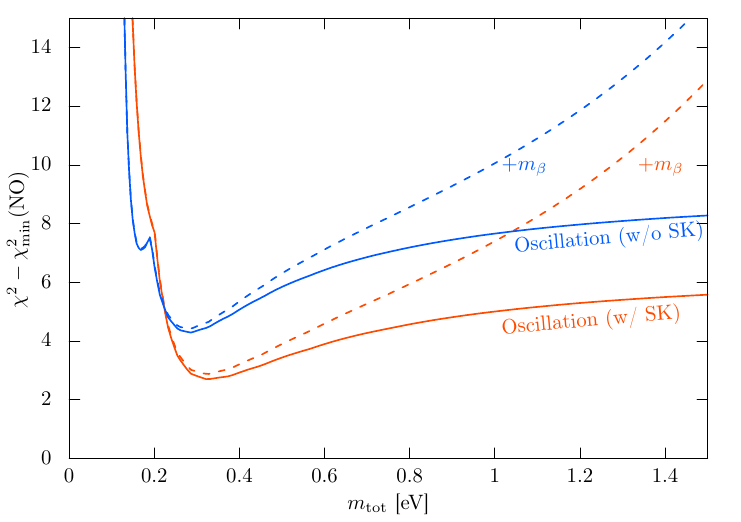}}
 	\subcaptionbox{IO}	{\includegraphics[width=0.49\textwidth]{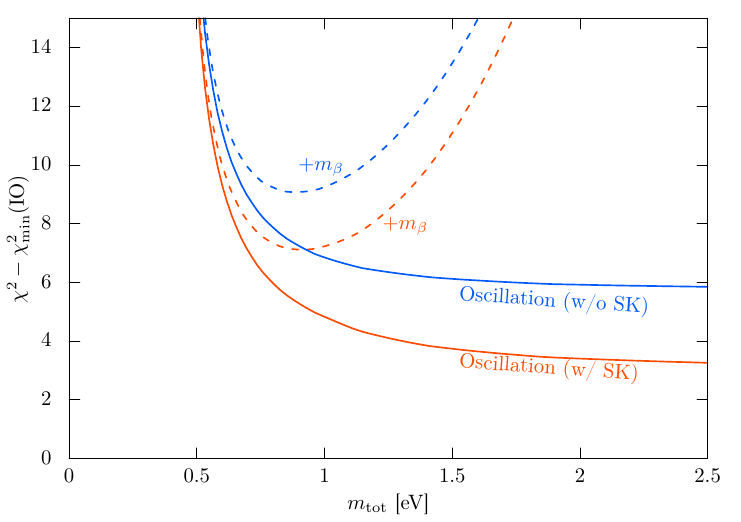}}
 \caption{
The difference in $\chi^2_\mathrm{eff}$ between two cases: one where the neutrino oscillation parameters are varied to minimize $\chi^2_\mathrm{eff}$ under the constraints of the minimal $\Umt$ model, and another where all six parameters are freely varied to minimize $\chi^2_\mathrm{eff}$. The latter is denoted as $\chi^2_\mathrm{min}$. Panel (a) corresponds to the NO case, while panel (b) corresponds to the IO case. The red (blue) curves are based on $\chi^2$ data from NuFIT~6.0 
``\texttt{IC24 without SK-atm}"
(``\texttt{IC19 w/o SK-atm}"). 
The dashed lines represent cases that, in addition to oscillation data, data from the direct neutrino mass measurement by KATRIN is included.
}
 \label{fig:chi2_osci}
\end{figure}

In Fig.\,\ref{fig:chi2_osci}, we plot, for each $m_\mathrm{tot}$, the difference in $\chi^2_\mathrm{eff}$ between two cases: 
one where the neutrino oscillation parameters are varied to minimize $\chi^2_\mathrm{eff}$ under the constraints of the minimal $\Umt$ model, 
and another where all six parameters are freely varied to minimize $\chi^2_\mathrm{eff}$. 
The latter is denoted as $\chi^2_\mathrm{min}$.
For the
\texttt{IC24 with SK-atm}
analysis,
we find that 
$\chi^2_\mathrm{min}=0$ for the NO case and $\chi^2_\mathrm{min} = 6.1$ for the IO case.
In contrast, 
for the 
\texttt{IC19 w/o SK-atm} analysis, 
$\chi^2_\mathrm{min}= 0.6$ for the NO case and $\chi^2_\mathrm{min}=0$ for the IO case (see Ref.\,\cite{Esteban:2024eli}).

It is important to note that the constraints in Eqs.\,\eqref{eq:r21} and \eqref{eq:r31} reduce the number of parameters 
by two, one of which corresponds to the absolute neutrino mass and is thus irrelevant for $\chi^2_\mathrm{eff}$. 
Consequently, due to Wilks' theorem,  the minimum value of the difference $\chi^2_\mathrm{eff} - \chi^2_\mathrm{min}$ follows a 
$\chi^2$ distribution with one degree of freedom.%
\footnote{Strictly speaking, this statement holds for $\chi^2_{\mathrm{full}}$. However, since $\chi^2_\mathrm{eff}$ provides a conservative estimate of $\chi^2_{\mathrm{full}}$, we assume that it also holds for $\chi^2_\mathrm{eff}$ in the following discussion.}
The minimum values of $\chi^2_\mathrm{eff}-\chi^2_{\min}$ 
are summarized in Tabs.\,\ref{tab:NO_summary} and \ref{tab:IO_summary}.
For the IO case, 
this minimum value is reached in the limit $m_\mathrm{tot}\to \infty$, 
where 
$\theta_{23}\to \pi/4$.

\subsection{Direct neutrino mass measurement}

The KATRIN experiment directly probes the absolute neutrino mass scale by analyzing the kinematics of tritium $\beta$-decay\,\cite{KATRIN:2005fny}. 
By examining the endpoint region of the electron energy spectrum, 
KATRIN constrains the effective electron neutrino mass, which is defined as  
\begin{equation}
\label{eq:mbeta}
    m_{\beta}^2 = {\sum_i |U_{ei}|^2 m_i^2} \ .
\end{equation}
Through a detailed spectral shape analysis near the endpoint, 
KATRIN places an upper bound on $m_{\beta}$, finding no significant deviation from the standard massless neutrino hypothesis\,\cite{Katrin:2024tvg},
\begin{equation}
    m_{\beta} < 0.45\,\text{eV}\,(90\% \,\text{CL})\ .
\end{equation}

\begin{figure}[t!]
	\centering
\includegraphics[width=0.5\textwidth]{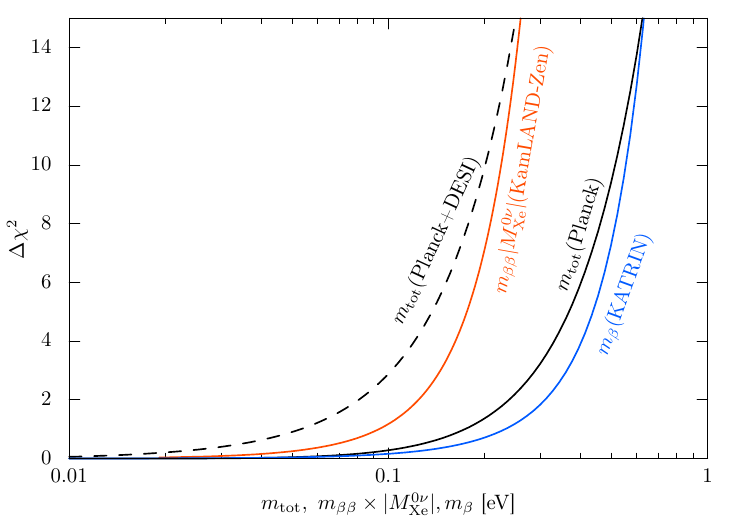}
\caption{
The $\Delta \chi^2$ values for the various neutrino mass parameters. 
The blue solid line represents $\Delta \chi^2$ from the direct mass measurement by KATRIN as a function of $m_\beta$ in Eq.\,\eqref{eq:mbeta}. 
The red solid line corresponds to $\Delta \chi^2$ from neutrinoless double-beta decay measured by KamLAND-ZEN as a function of $m_{\beta\beta} \times |M^{0 \nu}_{\mathrm{Xe}}|$ in Eq.\,\eqref{eq:mbetabeta}. 
The black solid (dashed) lines indicate $\Delta \chi^2$ from the cosmological observation of the total neutrino mass by Planck (Planck+DESI).
}
 \label{fig:mass}
\end{figure}

We extract the $\chi^2$ values for $m_{\beta}^2$ from Ref.\,\cite{Katrin:2024tvg} and fit them with a quadratic function.  
Figure\,\,\ref{fig:mass} shows $\Delta \chi^2$ as a function of $m_{\beta}$, with the blue line representing the fitted result.
The dashed lines in Fig.\,\ref{fig:chi2_osci} illustrate $\Delta \chi^2$ when KATRIN data is incorporated into the neutrino oscillation analysis.

Note that the estimation of $m_{\beta}$ requires knowledge of the full PMNS matrix and all three neutrino masses.  
In the minimal $\Umt$ model, the constraints in Eqs.\,\eqref{eq:r21} and \eqref{eq:r31} impose two relations among $m_1$, $m_2$, and $m_3$.  
As a result, when applying Wilks' theorem, the $\chi^2$ test must be performed with two degrees of freedom.

One important point should be emphasized here. 
For instance, when comparing the $\Delta\chi^2$ values between the analysis using only oscillation data in Tab.\,\ref{tab:NO_summary} and the analysis incorporating direct neutrino mass measurement results, the latter yields slightly larger values. 
However, as discussed above, the latter requires a $\chi^2$ analysis with two degrees of freedom. 
Consequently, the constraints obtained from this combined analysis are weaker than those derived solely from neutrino oscillation data.
This indicates that incorporating additional information into the analysis does not necessarily strengthen the constraints.

This can be understood as follows. 
Wilks' theorem assumes that the likelihood ratio test statistic follows a $\chi^2$ distribution, which in turn requires the Fisher information matrix to be well-conditioned.  
In the present case, some eigenvalues of the Fisher information matrix can be nearly zero, indicating that certain parameter directions are weakly constrained.
As a result, the likelihood function deviates from a quadratic form near its maximum, leading to a breakdown of the $\chi^2$ approximation underlying Wilks' theorem.
In such cases, relying solely on neutrino oscillation data provides a more conservative bound, as indicated by $\dagger$ in Tab.~\ref{tab:NO_summary}.

\subsection{Neutrinoless double-beta decay}

In the present model, 
neutrinos have Majorana masses, 
which leads to lepton number violating processes. 
The rates of these processes increase proportionally to the square of the neutrino mass. 
Currently, no lepton number violating processes have been observed experimentally, with the strongest constraints coming from neutrinoless double-beta ($0\nu\beta\beta$) decay in nuclei.

The decay half-life 
of a nucleus via the 
$(0\nu\beta\beta)$ decay, $T_{1/2}^{0\nu}$, 
is given by  
\begin{equation}
\label{eq:mbetabeta}
    \frac{1}{T_{1/2}^{0\nu}} = G^{0\nu} |M_{\mathrm{Xe}}^{0\nu}|^2 \frac{m_{\beta\beta}^2}{m_e^2} \ ,
\end{equation}
where $ G^{0\nu} $ is the phase-space factor, 
$ M_{\mathrm{Xe}}^{0\nu} $ is the nuclear matrix element of $^{136}$Xe, 
$ m_{\beta\beta} $ is the effective Majorana neutrino mass, 
and $ m_e $ is the electron mass (see  Ref.\,\cite{ParticleDataGroup:2024cfk} for a review). 
The effective Majorana neutrino mass is defined as  
\begin{equation}
    m_{\beta\beta} = \left| \sum_i U_{ei}^2 m_i \right| \, .
\end{equation}

\begin{figure}[t!]
	\centering
 	\subcaptionbox{NO}
{\includegraphics[width=0.49\textwidth]{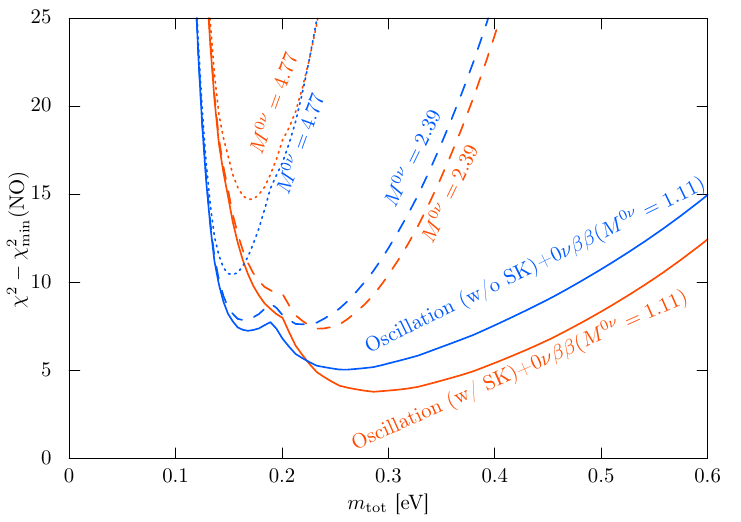}}
 	\subcaptionbox{IO}	{\includegraphics[width=0.49\textwidth]{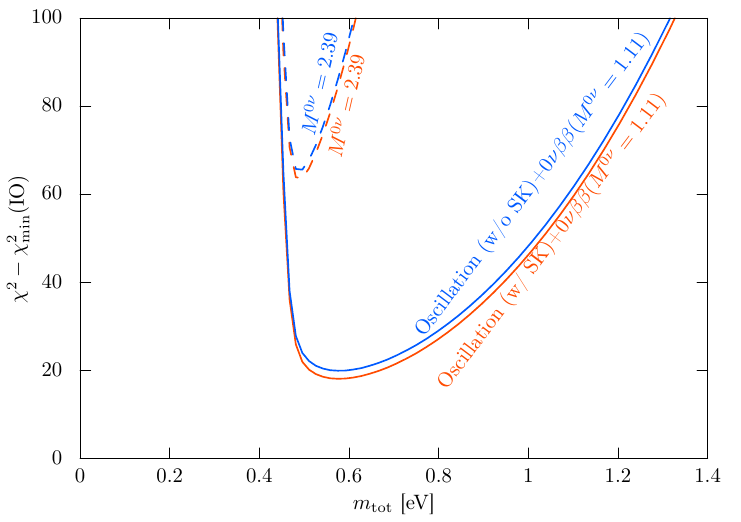}}
 \caption{
The values of $\Delta \chi^2$ incorporating both neutrino oscillation data and neutrinoless double-beta decay data from KamLAND-ZEN.
}
 \label{fig:0vchi}
\end{figure}

KamLAND-ZEN experiment searches for $(0\nu\beta\beta)$ decay using $^{136}$Xe, 
and so far, 
no signals have been detected,
which places
$T^{0\nu}_{1/2} > 3.8\times 10^{26}$\,yr at 90\%\,CL\,\cite{KamLAND-Zen:2024eml}.
In this paper, we obtain the $\chi^2$ for $ m_{\beta\beta} $ using data from KamLAND-ZEN.
We perform a fit for the background and signal in the observed spectrum within the energy range $ 2.35\,\mathrm{MeV} < E < 2.70$\,MeV. 
Considering the uncertainty in the nuclear matrix element,
the $\chi^2$ is evaluated for the product $ m_{\beta\beta} \times |M^{0\nu}_{\mathrm{Xe}}| $. 
In Fig.\,\ref{fig:mass}, we show the $\Delta \chi^2$ in red line.
Following the KamLAND-ZEN analysis, 
we analyze the range  
\begin{equation}
    1.11 \leq |M_{\mathrm{Xe}}^{0\nu}| \leq 4.77 \ , 
\end{equation}
for the nuclear matrix element uncertainty (see also e.g., Ref.\,\cite{Gomez-Cadenas:2023vca} for review).
Fig.\,\ref{fig:0vchi} presents $\Delta \chi^2$ as a function of $m_{\mathrm{tot}}$, 
incorporating information from neutrino oscillations and $m_{\beta \beta}$.  
For parameters other than $m_{\mathrm{tot}}$, 
the values that minimize $\Delta \chi^2$ are used, 
while multiple possibilities for the matrix elements are considered.

The estimation of $m_{\beta\beta}$ requires all neutrino parameters, including the Majorana phases.  
In the case of the minimal $\Umt$ model, the constraints given in Eqs.\,\eqref{eq:r21} and \eqref{eq:r31} reduce the number of free parameters by four in terms of real degrees of freedom. 
Accordingly, when applying Wilks' theorem, a $\chi^2$ test should be performed with four degrees of freedom.

\subsection{Cosmology}
Current upper limits on $m_\mathrm{tot}$
from cosmological observation
range from approximately $0.1$\,eV to $0.5$\,eV at 95\%\,CL, 
depending on the likelihood
profile, datasets, and the adopted cosmological model (see  Ref.\,\cite{ParticleDataGroup:2024cfk} for a review).
In the following analysis, 
we assume the $\Lambda$CDM model and use the likelihood profile provided in Ref.\,\cite{Naredo-Tuero:2024sgf}.
In particular, 
for the CMB constraints, 
we use the likelihood profile obtained from \texttt{HiLLiPoP23-PR4+Lensing-PR3}\,\cite{Tristram:2023haj,Tristram:2021tvh,Tristram:2020wbi,Planck:2018nkj}, 
where the known lensing anomaly present in the \texttt{Planck 2018 Plik} likelihood\,\cite{Planck:2018vyg} is significantly reduced.
For the BAO constraints, 
we adopt the \texttt{DESI-Y1-no07} dataset, 
which excludes the $z=0.7$ bin identified as a $2.6\sigma$ outlier in the DESI-Y1 results\,\cite{DESI:2024uvr,DESI:2024lzq}.
According to the analysis in Ref.\,\cite{Naredo-Tuero:2024sgf}, 
the combination of \texttt{HiLLiPoP23-PR4+Lensing-PR3} and \texttt{DESI-Y1-no07} provides the maximally conservative constraint 
from the combined CMB and BAO observations 
within the $\Lambda$CDM model.
In Fig.\,\ref{fig:mass}, 
we show those profile likelihoods.
\begin{figure}[t!]
	\centering
 	\subcaptionbox{NO}
{\includegraphics[width=0.49\textwidth]{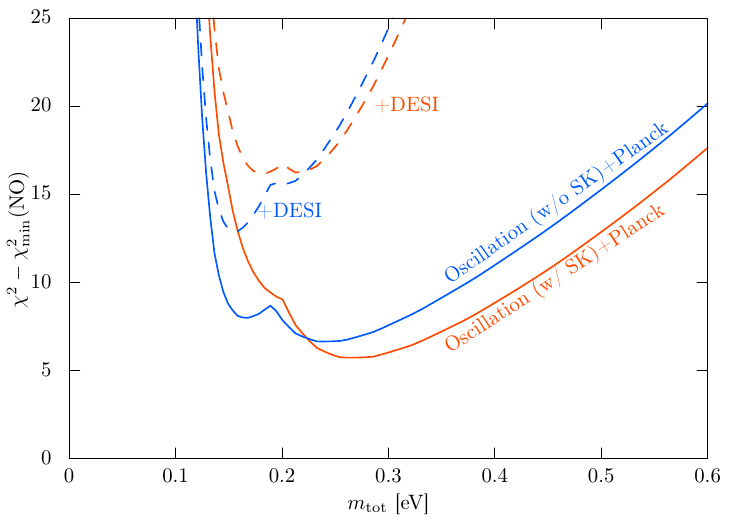}}
 	\subcaptionbox{IO}	{\includegraphics[width=0.49\textwidth]{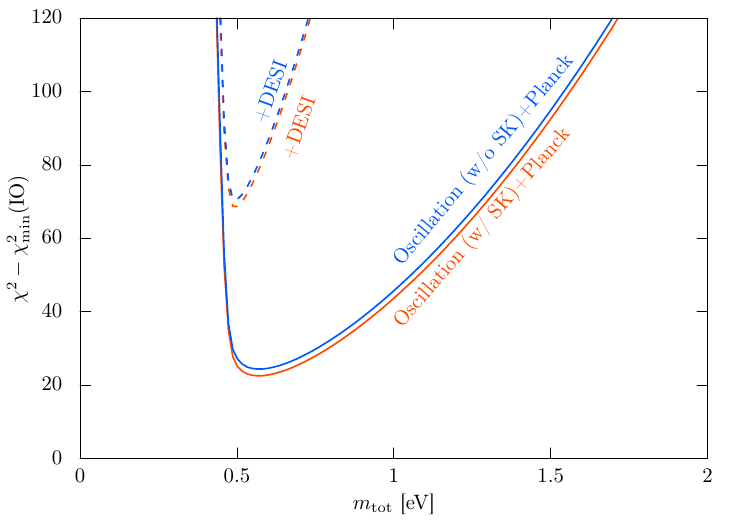}}
 \caption{
The $\Delta \chi^2$ values as a function of $m_{\mathrm{tot}}$. 
The blue (red) solid line represents $\Delta \chi^2$ obtained using the NuFIT 6.0 data without (with) SK consideration.
The cosmological data used are from Planck and DESI.
}
 \label{fig:chi2_cosmo}
\end{figure}
Figure\,\,\ref{fig:chi2_cosmo} shows $\Delta \chi^2$ as a function of $m_{\mathrm{tot}}$, 
incorporating information from neutrino oscillations and cosmology.
Note that as in the case of the $\chi^2$ analysis with the direct neutrino mass measurement,  
the $\chi^2$ test must be performed with two degrees of freedom.

\section{Conclusion}
\label{sec:conclusion}

In this work, we examined the minimal $\Umt$ gauge model in light of the latest neutrino data, incorporating results from neutrino oscillation experiments, cosmological observations, direct neutrino mass measurements, and searches for neutrinoless double-beta decay. Our analysis was conducted within a frequentist framework to ensure robustness against prior-dependent uncertainties.

Using the most conservative neutrino oscillation data, we found that normal ordering (NO) is excluded at approximately the 90\% confidence level (CL). Incorporating cosmological constraints from the Planck 2020 PR4 analysis strengthens this exclusion to about 95\%\,CL, 
while further including BAO data increases it to nearly 99\%\,CL. 
For the inverted ordering (IO) case, 
the constraints are even more severe, 
with oscillation data alone excluding it at the 92\%\,CL, 
and additional cosmological constraints leading to a statistical significance of $4.4\sigma$. 

The primary reason for this exclusion is the model’s predictive structure, which leads to a nearly degenerate neutrino mass spectrum. This results in a large total neutrino mass, which is strongly constrained by cosmological data. Additionally, neutrinoless double-beta decay and direct neutrino mass measurements impose further restrictions on the parameter space.

Our results indicate that the minimal $\Umt$ gauge model is increasingly disfavored under the assumption of the $\Lambda$CDM framework. Reviving this model would require significant modifications, 
such as introducing additional fields that alters the two-zero minor structure of the neutrino mass matrix or relax the stringent cosmological constraints.

Future advancements in neutrino physics, particularly improved measurements of the absolute neutrino mass scale and cosmological constraints, will further test the viability of this model. If stronger constraints on the total neutrino mass emerge, the minimal $\Umt$ scenario may be entirely ruled out, necessitating a reconsideration of its role in particle physics.

In this work, we have primarily focused on the most conventional realization of the seesaw mechanism, adopting it as a minimal model. Here, we briefly comment on alternative neutrino mass generation mechanisms.
If neutrinos are Dirac fermions, the presence of ${\Umt}$ symmetry enforces a diagonal mass matrix at the renormalizable level, rendering the observed neutrino mixing angles unexplained within such a framework.
For the type-III seesaw mechanism \cite{Foot:1988aq}, where fermionic SU(2) triplets are introduced to generate neutrino masses, the arguments presented in this paper apply straightforwardly, leading to the same constraints on the neutrino mass matrix.
In contrast, the minimal type-II seesaw model \cite{Magg:1980ut,Cheng:1980qt,Lazarides:1980nt,Mohapatra:1980yp}, where an SU(2) triplet scalar is responsible for generating neutrino masses, is not consistent with current experimental data under ${\Umt}$ symmetry. However, by extending the model to include multiple SU(2) triplet scalars that transform appropriately under ${\Umt}$, it is possible to obtain a richer structure for the neutrino mass matrix.
An analysis of such extended scenarios is presented in the Appendix.

In this analysis, we have neglected finite threshold corrections from the $\Umt$ gauge symmetry breaking sector. However, if the relevant couplings are sufficiently strong, these corrections could disrupt the two-zero minor structure of the neutrino mass matrix. A more detailed investigation of these effects would be worthwhile.

\section*{Acknowledgements}
This work is supported by Grant-in-Aid for Scientific Research from the Ministry of Education, Culture, Sports, Science, and Technology (MEXT), Japan, 21H04471, 22K03615, 24K23938 (M.I.), 20H01895 and 20H05860  (S.S.) and by World Premier International Research Center Initiative (WPI), MEXT, Japan. 
This work is supported by JST SPRING, Grant Number JPMJSP2108 and ANRI fellowship (K.W.).

\appendix

\section{Constraints on Two-Zero Textures in \texorpdfstring{$\boldsymbol{\Umt}$}{} Model}
\begin{figure}[t!]
	\centering
 	\subcaptionbox{NO}
{\includegraphics[width=0.49\textwidth]{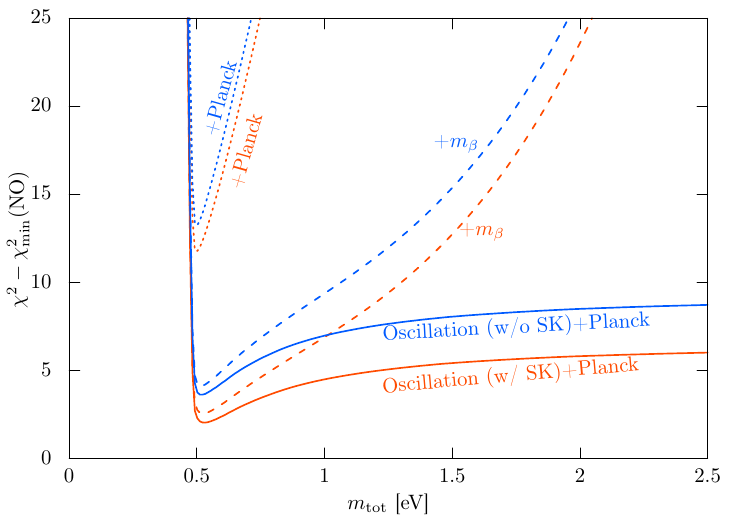}}
 	\subcaptionbox{IO}	{\includegraphics[width=0.49\textwidth]{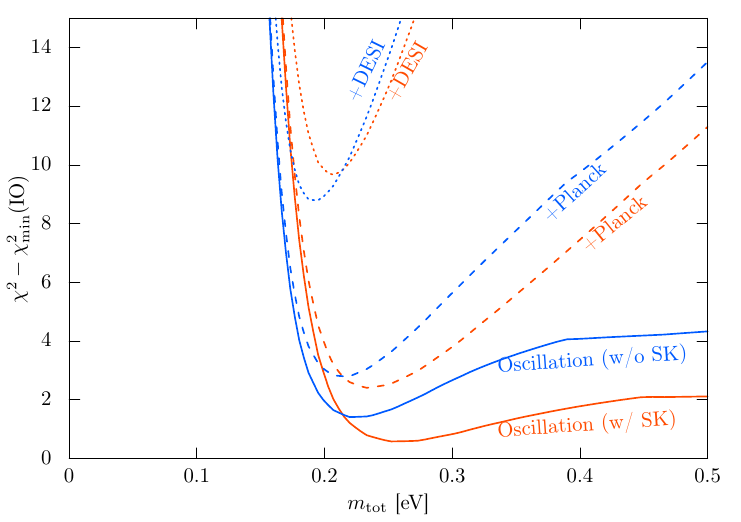}}
 \caption{
The values of $\Delta \chi^2$ incorporating neutrino oscillation data, cosmological observations, and direct neutrino mass measurements, under the condition $(m_\nu)_{\mu\mu} = (m_\nu)_{\tau\tau} = 0$.
}
 \label{fig:major}
\end{figure}

In the minimal type-I seesaw model, the ${\Umt}$  symmetry imposes a constraint on the inverse neutrino mass matrix, requiring that $m_{\nu}^{-1}$  has vanishing  $(\mu,\mu)$ and $(\tau,\tau)$ entries.
In contrast, in several non-minimal neutrino mass generation scenarios, the same ${\Umt}$ symmetry can directly constrain the neutrino mass matrix itself, leading to 
\begin{align}
    (m_{\nu})_{\mu\mu} = (m_{\nu})_{\tau\tau} = 0\, .
\end{align}
Such relations can arise, for example, in type-II seesaw models where three SU(2) triplet scalars with ${\Umt}$ charges 0, $+1$, and $-1$ are introduced to generate the neutrino mass matrix. 
Similar structures can also be realized in radiative neutrino mass generation mechanisms \cite{Baek:2015mna,Lee:2017ekw} and in inverse seesaw models \cite{Dev:2017fdz}.
 This condition again imposes relations among $m_1$, $m_2$, and $m_3$, allowing for statistical analyses analogous to those presented in the main text.
Figure \ref{fig:major} shows the $\Delta \chi^2$ as a function of the total neutrino mass $m_\mathrm{tot}$ for the NO and IO cases.
In the NO case, neutrino oscillation data imposes a lower bound of $m_\mathrm{tot} \gtrsim 0.5$ eV, which is in significant tension with cosmological constraints.
For the IO case, this lower bound is relaxed to $m_\mathrm{tot} \gtrsim 0.2$ eV, resulting in a milder, though still non-negligible—conflict with cosmological observations, particularly with BAO data.
The resulting constraints for both cases are summarized in Tables \ref{tab:major_NO_summary} and \ref{tab:major_IO_summary}.
The degrees of freedom used in the $\chi^2$ analysis for each dataset are the same as those described in the main text.

\begin{table}[h]
    \centering
        \caption{Constraints on neutrino properties under the two-zero texture condition $(m_\nu)_{\mu\mu} = (m_\nu)_{\tau\tau} = 0$ in the NO case. }
\label{tab:major_NO_summary}
    \begin{tabular}{c|c|c|c|c} \hline
        Oscillation & $m_{\beta}$ & Cosmology & $\Delta \chi^2$(NO) & Confidence Level \\ \hline
        \multicolumn{5}{c}{\textbf{IC19 w/o SK-atm (NuFIT 6.0)}} \\ \hline
        NuFIT 6.0 & - & - & $3.6$ & $94\%$\,CL$^{\phantom{\dagger}}$ \\
        NuFIT 6.0 & KATRIN  & - & $4.1$ & $94\%$\,CL$^\dagger$ \\ \hline
        NuFIT 6.0 & - & Planck & $13$ & $3.2\sigma$ \\
        NuFIT 6.0  & - & Planck + DESI & $57$ & $7.2\sigma$  \\ \hline
        \multicolumn{5}{c}{\textbf{IC24 with SK-atm (NuFIT 6.0)}} \\ \hline
        NuFIT 6.0 & - & - & $2.0$ & $84\%$\,CL$^{\phantom{\dagger}}$ \\
        NuFIT 6.0 & KATRIN & - & $2.6$ & $84\%$\,CL$^\dagger$ \\ \hline
        NuFIT 6.0 & - & Planck & $12$ & $3.0\sigma$  \\
        NuFIT 6.0 & - & Planck + DESI & $55$ & $7.1\sigma$  \\ \hline
    \end{tabular}
\end{table}

\begin{table}[h]
    \centering
    \caption{Same as Tab.\,\ref{tab:major_NO_summary}, but for the IO case.}
    \label{tab:major_IO_summary}
    \begin{tabular}{c|c|c|c|c} \hline
        Oscillation & $m_{\beta}$ & Cosmology & $\Delta \chi^2$(IO) & Confidence Level \\ \hline
        \multicolumn{5}{c}{\textbf{IC19 w/o SK-atm (NuFIT 6.0)}} \\ \hline
        NuFIT 6.0 & - & - & $1.4$ & $76\%$\,CL$^{\phantom{\dagger}}$ \\
        NuFIT 6.0 & KATRIN  & - & $1.5$ & $76\%$\,CL$^\dagger$ \\ \hline
        NuFIT 6.0 & - & Planck & $2.9$ & $77\%$\,CL$^{\phantom{\dagger}}$ \\
        NuFIT 6.0  & - & Planck + DESI & $8.9$ & $2.5\sigma$  \\ \hline
        \multicolumn{5}{c}{\textbf{IC24 with SK-atm (NuFIT 6.0)}} \\ \hline
        NuFIT 6.0 & - & - & $0.6$ & $56\%$\,CL$^{\phantom{\dagger}}$ \\
        NuFIT 6.0 & KATRIN & - & $0.7$ & $56\%$\,CL$^\dagger$ \\ \hline
        NuFIT 6.0 & - & Planck & $2.4$ & $70\%$\,CL$^{\phantom{\dagger}}$   \\
        NuFIT 6.0 & - & Planck + DESI & $9.7$ & $2.6\sigma$  \\ \hline
    \end{tabular}
\end{table}

\bibliographystyle{apsrev4-1}
\bibliography{bibtex}

\end{document}